\documentclass[preprint2]{aastex}

\shorttitle{Quiescent observations of PQ And}
\shortauthors{Schwarz et al.}

\begin{document}

\title{Quiescent observations of the WZ Sge type dwarf nova PQ Andromedae}

\author{Greg J. Schwarz\altaffilmark{1}, 
Travis Barman\altaffilmark{2}, 
Nicole Silvestri\altaffilmark{3}, 
Paula Szkody\altaffilmark{3}, 
Sumner Starrfield\altaffilmark{4}, 
Karen Vanlandingham\altaffilmark{1,5}, and
R. Mark Wagner\altaffilmark{6}
}

\altaffiltext{1}{Steward Observatory, 933 N. Cherry Ave., 
Tucson, AZ 85721, gschwarz@as.arizona.edu}
\altaffiltext{2}{Dept. of Physics, Wichita State University, 
1845 Fairmount Rd., Box 32, Wichita, KS 67260-0032}
\altaffiltext{3}{Dept. of Astronomy, University of Washington, Box 351580, 
Seattle, WA 98195}
\altaffiltext{4}{Dept. of Physics and Astronomy, Arizona State University, 
Tempe, AZ 85287-1504}
\altaffiltext{5}{Dept. of Astronomy, Columbia University, 1328 Pupin Hall,
550 W. 120th Street, New York, NY 10027}
\altaffiltext{6}{LBT Observatory, University of Arizona, 933 N. Cherry Ave.,
Tucson, AZ 85721} 

\begin{abstract}

We have obtained time series optical spectra of the cataclysmic variable
PQ And in quiescence.  The spectra show a white dwarf continuum with
narrow Balmer emission superimposed over strong Balmer absorption.  The
emission lines have blue and red components whose strength changes
with time.  An analysis of the H$\alpha$ emission line implies a short
orbital period below the period gap.  Given its lack of accretion disk
features, its large and infrequent outbursts, and an orbital period below
the period gap, PQ And is probably a low accretion rate object similar
to WZ Sge.  In addition, white dwarf model fits imply that PQ And
is an excellent ZZ Cet candidate.

\end{abstract}


\section{Introduction}

PQ And was discovered by McAdam on 21 March 1988 \citep{iauc4570} at 
a visual magnitude of 10.  Examination of the Palomar Sky Survey plates 
showed a precursor object with a blue magnitude between 18-19 and a 
red magnitude of about 20 \citep{iauc4577,iauc4579}.  Within 19 days 
the light curve had declined by 2 magnitudes.  The large outburst 
amplitude and rapid decline led to the initial classification of a 
classical nova but spectra taken 3.5 months after the outburst by 
\citet{iauc4629} lacked the nebular features typically observed in 
classical novae.  Instead, the spectra had strong Balmer emission 
surrounded by broad absorption similar to the quiesent spectra of 
the dwarf nova WZ Sge.  The strong absorption indicated that the 
energy distribution was primarily from the white dwarf (WD) and implied 
an extremely low accretion rate. Recently, it has been shown that 
systems with this type of spectra are good candidates for containing 
non-radially pulsating DA WDs or ZZ Cet stars \citep{WW04a,WW04b}.

WZ Sge is a dwarf nova whose outbursts occur on timescales of tens of
years and with amplitudes of 7-8 magnitudes.  Cataclysmic variables (CVs) 
of this type are also known as TOADS or Tremendous Outburst Amplitude 
Dwarf novae \citep{HSC95}.  These characteristics are thought to be 
due to the very close binary separation and the extremely low accretion 
rate in these systems.  TOAD orbital periods are all below the period 
gap and are typically of order 90 minutes.  The mass transfer rates 
implied by the models are only 10$^{-11}$M$_{\odot}$ yr$^{-1}$ 
\citep{HSC95}.

Given PQ And's similarities to a dwarf nova, \citet{richter90} searched 
archival plates for evidence of previous outbursts.  Two other outbursts 
with maximum magnitudes of $\sim$ 11 were found on 23 August 1938 
and 7 March 1967.  With decades long superoutburst timescales, an
outburst amplitude of $\sim$ 9 mag, and a WZ Sge like quiescent spectrum, 
PQ And clearly has TOAD characteristics.  In this paper we show 
that the orbital period {\em appears to be} below the period gap 
and that the system likely contains a WD in the instability strip.

\section{Observations}

We obtained 10 spectra of PQ And on the 6.5m Multiple Mirror 
Telescope (MMT) on 16 September 2003.  We used the blue channel CCD 
spectrograph with a 500 grooves per millimeter grating and a 1\arcsec\ 
wide slit for a 
spectral resolution of 3.6\AA.  Spectral coverage extends from 
4100\AA\ to 7200\AA.  The exposure 
time was set to 480 seconds in order to resolve an orbital period as 
low as 80 minutes.  The seeing ranged from 1\farcs0-1\farcs5 
during the night.  Biases and flat fields were obtained at the beginning 
and end of each night and a HeNeAr comparison lamp spectrum was 
obtained after each PQ And observation for optimum wavelength 
calibration.  Standard stars were also observed throughout 
the night for flux calibration.  The spectra were reduced using 
standard IRAF\footnote{ IRAF is distributed by the National 
Optical Astronomy Observatories, which are operated by the Association 
of Universities for Research in Astronomy, Inc., under cooperative 
agreement with the National Science Foundation.} routines.

\begin{figure}
\plotone{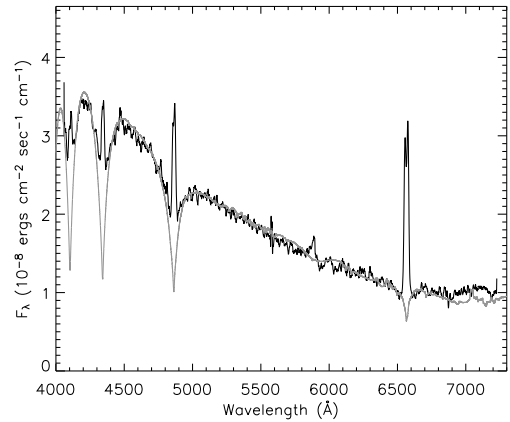}
\caption{The combined spectrum (black line) and the
best fit {\tt PHOENIX} white dwarf model (grey line).  The white dwarf 
model parameters are T$_{eff}$ = 12,000 K, log($g$) = 7.7, 
Z = 10$^{-2}$ Z$_{\sun}$.  The model also includes the flux from a 
M dwarf secondary with a T$_{eff}$ = 2500 K.}
\label{modelfit}
\end{figure}

\begin{figure}
\plotone{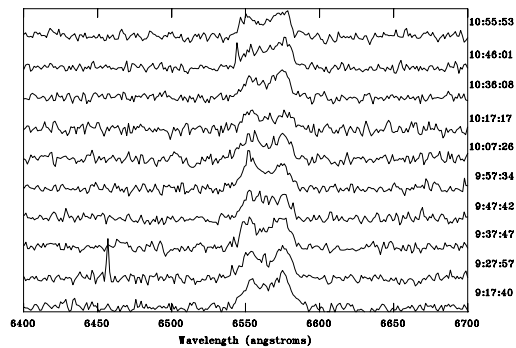}
\caption{The evolution of H$\alpha$. The numbers on the right
hand side of the figure give the time of mid exposure for each
spectrum.  NOTE TO EDITOR: This figure requires
a minimum of 1/2 a page to show its details.}
\label{evolve}
\end{figure}

The combined spectrum is shown in Figure 
\ref{modelfit}\footnote{Since the extinction toward PQ And provided 
by the \citet{SFD98} maps is extremely low, E($B-V$) = 0.06, none of 
the spectra shown in this paper have been dereddened nor have they
been Doppler corrected.}.  The combined 
spectrum has narrow emission lines flanked by broad absorption similar 
to that described by \citet{iauc4629}.  The equivalent widths, fluxes, 
and FWZI values of the observed emission lines in the combined 
spectrum are given in Table \ref{tab:emission}.  All the emission 
lines have a red and a blue component whose strength and peak vary 
with time.  The evolution of these components can best be seen in 
H$\alpha$ and is shown in Figure \ref{evolve}.  

\section{Emission line analysis}

Radial velocities from the line wings were measured by convolving 
the line profiles with a pair of Gaussian bandpasses 
\citep[see][for details]{shaft,P1} whose width is set to the resolution 
and whose separation is varied.  For each separation, the velocities are 
fit to a sine function of the form 
\begin{equation}
v = \gamma - K\rm sin[2\pi(\phi - \phi_{0})]
\end{equation}
where $\gamma$ is the systemic velocity, K is the semiamplitude, 
and $\phi$ is the offset from superior conjunction of the WD.  The 
best solution is chosen as the one producing the smallest errors
in the individual parameters and the total $\sigma$ of the fit.

\begin{figure} 
\plotone{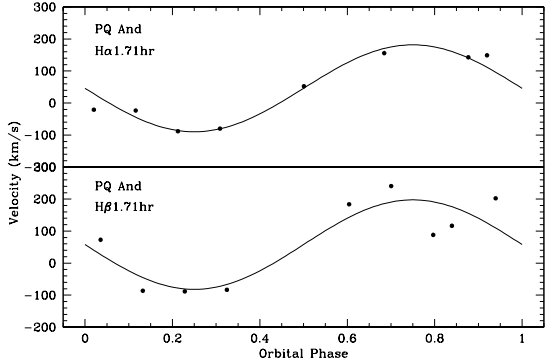} 
\caption{Radial velocity data obtained from a double Gaussian 
fitting to the line wings.  Top panel gives the radial velocities as 
determined from the H$\alpha$ emission feature.  The bottom panel gives 
the radial velocities as determined from the H$\beta$ emission feature.  
The best-fit solution yields a period of 1.7 hours.} 
\label{radvel} 
\end{figure}

We fit both the H$\alpha$ and H$\beta$ lines with this method but
unfortunately not all the observed line profiles were useful.
Two H$\alpha$ and one H$\beta$ profiles were discarded due to 
cosmic rays or noise at a particular separation.  Even with one 
less data point the H$\alpha$ fit was significantly better than 
that determined for H$\beta$ thus we adopt the H$\alpha$ separation 
and period as the best solution and force H$\beta$ to match.  
The best fit period is 1.7$\pm$0.1 hours where this is the formal
uncertainty from the analysis.  The true uncertainty in the period is
greater since the derived period is longer than the sampling time.  
Figure \ref{radvel} shows the radial velocity data for H$\alpha$ and 
H$\beta$ fits.  The radial velocity data is provided in Table \ref{tab:RV}.
The orbital parameters of the lowest total $\sigma$ fit are 
given in Table \ref{tab:parms} along with other parameters of PQ And.

\begin{figure}
\plotone{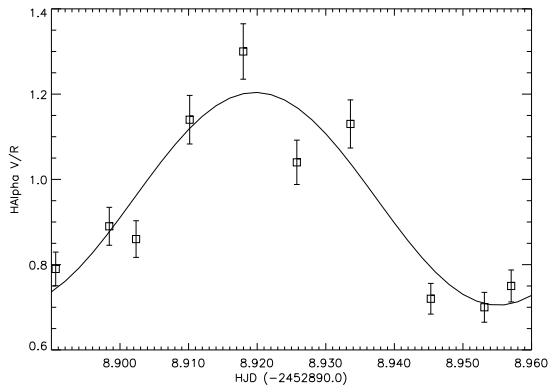}
\caption{The ratio of the blue (violet) peak to red H$\alpha$ peaks.
The solid line is the best sinusoidal fit assuming P$_{orb}$ = 1.7 h.}
\label{ratio}
\end{figure}

To quantify the component changes, we calculated the ratio of the blue
(violet) to red components (V/R) for each spectrum that shows H$\alpha$.
The V/R ratios were calculated with the IRAF program splot with two
Gaussians which had the same FWHM and were centered at 6554\AA and 6574\AA.  
In each spectrum the Gaussians were only allowed a single wavelength shift 
provided that the separation was always 20\AA.  The results are shown in 
Figure \ref{ratio}.

Superior conjunction of the WD is defined as the red to blue crossing 
point of the emission lines.  In Figure \ref{ratio} we have fit the 
V/R data with a sine wave of the form:
\begin{equation}
V/R = A + B\rm sin[2\pi(t-t_0/P_{orb})]
\end{equation}
to determine the crossing point.  With P$_{orb}$ = 1.7 hours the best fit 
gives a superior conjunction date of HJD 2,452,898.902.  The other model 
parameters are provided in Table \ref{tab:parms}.

\subsection{Differential photometry}

\begin{figure}
\plotone{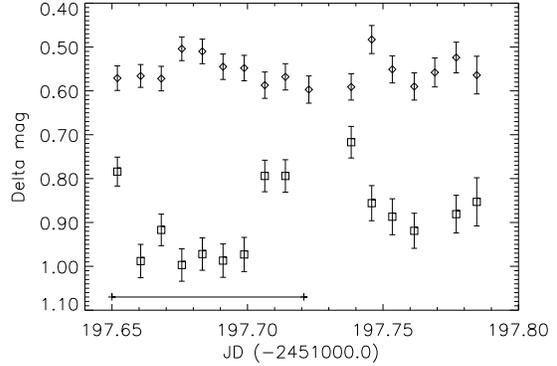}
\caption{The 19 January 1999 differential light curve.  The squares and 
diamonds are PQ And minus the comparison star and the comparison star minus 
the check star (plus 1 magnitude) values plus their summed photometric 
errors, respectively.  The line at the bottom left of the figure displays 
1.7 hours in this plot.}
\label{lightcurve}
\end{figure}

As an added check on the derived period we utilized differential photometry 
of PQ And from a CV orbital period survey taken in early 1999.  
The data were obtained at the Steward Observatory 1.6m Kuiper Telescope 
on Mount Bigalow with the 2K$\times$2K back-illuminated CCD Imager.  
The CCD was set at 2$\times$2 binning 
giving an image scale of 0.3$\arcsec$/pixel. The seeing varied during 
the run between 1$\arcsec$\ and 2$\arcsec$.  The observations were carried 
out in the $I$ band where the accretion disk and secondary star are 
typically of comparable brightness.  The exposure time in the PQ And
field was 600 seconds.  Biases and flats were taken at 
the beginning and end of each night and the data were reduced in IRAF.
The only night that we spent sufficient time to confirm a 1.7 hour
period was 19 January 1999.  On that night PQ And was observed 18 times
over 3 hours.  A differential light curve of the results is shown in
Figure \ref{lightcurve}.  The observed changes are suggestive of a
variation similar to our derived spectroscopic period.  The normalized
periodogram \citep{Scar82,HB86} of the photometry gives a longer period
of 2.1 hours but at a low confidence level of only 50\%.

We emphasize that these are not definitive orbital period solutions
since the derived periods are similar in length to the total time that
PQ And was observed.  More data are necessary to reach a more definitive
orbital period for this object.  However, the fact that the radial velocities
and the changes in the blue and red components show sine wave like behavior on
the timescale of our observation runs imply that the period must be short.
A firm upper limit on the orbital period can be established since the 
accretion disk would overwhelm the WD at periods $>$ 3 hours and there 
are no known dwarf novae in the gap.  Thus for the remainder of the paper 
we will assume an orbital period range between 1.7 and 2.1 hours.

\section{White dwarf model fitting}

\begin{figure}
\plotone{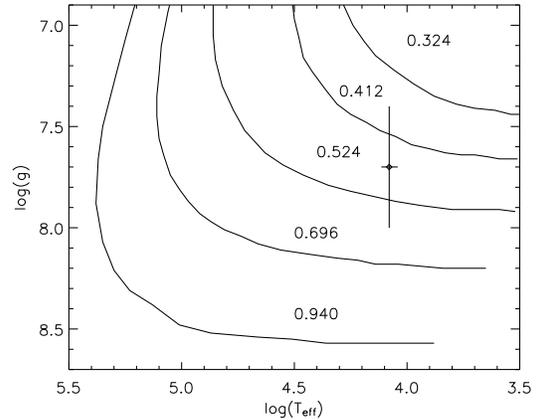}
\caption{The derived PQ And parameters on the theoretical white dwarf 
cooling tracks of \citet{Driebe98}. The White Dwarf mass is estimated
to be $\approx$ 0.5 $M_\odot$.}
\label{wdplot}
\end{figure}

The WD effective temperature and gravity were determined by comparing the
observed spectrum to a grid of synthetic spectra computed using the {\tt
PHOENIX} model atmosphere code.  Details of the modeling can be found in
\cite{Barman00}.  The grid includes models with $T_{eff}$ = 9,000K -- 15,000K
(in steps of 1000K), $\log(g)$ = 6.5, 7.0, 7.5, 8.0, and 10$^{-1}$, 10$^{-2}$
and solar metallicities.  After masking the Balmer emission lines due to the
accretion disk, $T_{eff}$ was first determined using least-squares minimization
across the full observed spectral range.  Fixing $T_{eff}$, the surface
gravity was found by fitting only the wings of H$\delta$, H$\gamma$, and
H$\beta$ since these were the only features available that are reasonably
sensitive to gravity.  The models with sub-solar metallicities gave only
marginally better fits; however, models with solar metallicities indicate 
that several metal absorption lines should have been detected but clearly 
were not observed.  Consequently, the metallicity is likely less than 
10$^{-1}$ solar.   Overall, varying metallicity had little effect on the 
final $T_{eff}$ and $\log(g)$ values.  The final best-fitting model had 
$T_{eff}$ = 12,000K $\pm$ 1000K and $\log(g)$ = 7.7 $\pm$ 0.3 (cgs units) 
with 10$^{-2}$ solar metallicity.  Estimates for the WD mass and radius 
were found by using theoretical cooling tracks on a $T_{eff}$ -- $\log(g)$ 
plane \cite[]{Driebe98}.  Figure \ref{wdplot} shows that the best fit 
$T_{eff}$ and $\log(g)$ values give $M_{wd} = 0.47\pm0.13 M_\odot$ and 
$R_{WD} = 0.0165 R_\odot$.  Note that this is only a mass estimate since 
the WDs in CVs are heated by accretion.  The WD in PQ And might not 
follow this exactly.

There was residual flux red-ward of 6000\AA\ between the observed and the
best fit WD model spectra.  This red excess is from the secondary.  To 
improve the model fit, a synthetic spectrum of an M dwarf was added to 
the best-fitting WD synthetic spectrum.  Since the secondary in CVs fill 
their Roche lobe, their characteristics are determined by the properties 
of the binary system.  Equation 2.101 in \citet{War95},
\begin{equation}
R_2(\odot) = 0.094 P^{1.08}_{orb}(h),
\end{equation}
relates the orbital period and secondary radius.  
Applying this equation to our derived periods gives a range of
$R_{2} = 0.17-0.20 R_\odot$.  The best fit effective temperature of the
secondary from visual inspection was 2500 K, however, this should not
be taken as definitive since the uncertainty is easily $\pm$ 500 K.
Even this range and the short orbital period still imply a late M dwarf 
as the secondary star.  The best fit WD plus secondary model 
spectral energy distribution is shown in Figure \ref{modelfit}.
An attempt was made to extract the secondary spectrum by subtracting 
the best-fitting WD synthetic spectrum from the observed spectrum.  
Unfortunately, the resulting spectrum was too noisy and did not cover 
enough of the TiO and VO bands to make a reliable determination of 
the spectral type.  The only reported infrared color for
PQ And, ($J-K$) = 0.85$\pm$0.49 \citep{SHM96}, is consistent within
the uncertainty with a late M spectral type.  PQ And is too faint
have been detected by the 2MASS survey \citep{2MASS}.

The mass equation, Equation 2.100 in \citet{War95},
\begin{equation}
M_2(\odot) = 0.065 P^{5/4}_{orb}(h),
\end{equation}
is valid for P$_{orb}$ between 1.3 and 9 hours and provides 
a secondary mass range of 0.13-0.16 M$_{\odot}$ for periods 
between 1.7 and 2.1 hours.  Given the derived WD mass the binary 
system mass ratio, $q = M_2/M_1$, is 0.28-0.34.

Using the derived WD radius, an absolute magnitude was computed for the
best-fitting WD synthetic spectrum; M$_v$ = 11.2.  With an apparent 
magnitude, m$_v$ = 19 \citep{DS93} and E(B-V) = 0.06, the distance to PQ And 
was found to be $330 \pm 50$ pc.  This distance places PQ And 110 pc below
the galactic plane.  The only other distance estimate in 
the literature is from \citet{SHM96}.  They use the observed $K$ band 
magnitude to determine the absolute magnitudes.  Their method depends 
on knowledge of the secondary's size and its fractional contribution to 
the $K$ band flux.  Without any period information \citet{SHM96} obtained 
a large range in distance for PQ And by assuming orbital periods between 
80 minutes and 6 hours.  Using their same method but now with our derived
period range gives a distance between 319-817 pc.  This estimate 
is still only an lower limit since it assumes that the secondary contributes 
100\% of the $K$ band flux however it is consistent with the distance 
from our WD analysis.  

The best fit {\tt PHOENIX} WD model puts PQ And in the region of
the ZZ Ceti instability strip \citep{B95}.  The low accretion rate implied
by the strong Balmer absorption makes PQ And an ideal candidate for
follow-up observations for non-radial pulsations.  If found, an analysis
of the oscillations will provide information on important WD parameters
including mass, composition, magnetic field, rotation, temperature, and
luminosity \citep{Win91}.  GW Lib \citep{van04}, SDSS J161033.64-010223.3
\citep{WW04a}, SDSSJ013132.39-090122.3 and SDSSJ220553.98+115553.7
\citep{WW04b} are other WZ Sge type dwarf novae of similar WD temperature
that have already been shown to exhibit non-radial pulsations.

\section{Conclusions}

PQ And has characteristics similar to WZ Sge type variable stars.  Its
quiescent spectra show little indication of an accretion disk
implying a low mass transfer rate.  PQ And had three known 
superoutbursts of amplitude $\sim$ 9 magnitudes in the 50 years
prior to 1988.  Fits to spectra give P$_{orb}$ = 1.7 hours while 
the period derived from photometry is slightly longer at 2.1 hours.  
More observations are required to confirm the true period but the 
available evidence strongly supports an orbital period below the orbital gap.
The best fit WD model gives T$_{eff}$ = 12,000 K,
log(g) = 7.7, and a distance of 330 pc.  The WD parameters place it
within the ZZ Cet instability strip.  With a quiescent magnitude of $V$ = 19, 
non-radial pulsations could be detected in PQ And with a mid-sized
telescope.

\acknowledgments

SS is grateful to the NSF and NASA for partial support of this research.


\clearpage

\begin{deluxetable}{lrcc}
\tablewidth{0pt}
\tablecaption{Emission line measurements\label{tab:emission}}
\tablehead{
\colhead{Line} & \colhead{EW} & \colhead{Flux} & \colhead{FWZI} \\
\colhead{} & \colhead{(\AA)} & \colhead{(erg s$^{-1}$ cm$^{-2}$)} &
\colhead{(km s$^{-1}$)}
}
\startdata
H$\delta$ & 4 & 2.6e-16 & 2100 \\
H$\gamma$ & 7 & 4.6e-16 & 2100 \\
H$\beta$ & 21 & 1.0e-15 & 2500 \\
\ion{He}{1} & 5 & 2.1e-16 & \nodata \\
H$\alpha$ & 75 & 2.1e-15 & 2700 \\
\enddata
\end{deluxetable}

\begin{deluxetable}{crrrr}
\tablewidth{0pt}
\tablecaption{Radial Velocity measurements\label{tab:RV}}
\tablehead{
\colhead{} & \multicolumn{2}{c}{H$\alpha$} &
\multicolumn{2}{c}{H$\beta$} \\
\colhead{HJD\tablenotemark{a}} & \colhead{Phase} & \colhead{RV} &
\colhead{Phase} & \colhead{RV} \\
\colhead{} & \colhead{} & \colhead{(km s$^{-1}$)} &
\colhead{} & \colhead{(km s$^{-1}$)}
}
\startdata
2898.890 & 0.92 & 149.1 & 0.84 & 116.6 \\
2898.897 & 0.02 & -20.9 & 0.94 & 202.4 \\
2898.904 & 0.12 & -23.4 & 0.04 &  72.8 \\
2898.911 & 0.21 & -88.1 & 0.13 & -86.4 \\
2898.918 & 0.31 & -80.0 & 0.23 & -88.2 \\
2898.925 & \nodata & \nodata & 0.32 & -83.3 \\
2898.932 & 0.50 &  52.3 & \nodata & \nodata \\
2898.945 & 0.68 & 155.9 & 0.60 & 183.9 \\
2898.952 & \nodata & \nodata & 0.70 & 240.7 \\
2898.958 & 0.88 & 142.7 & 0.80 & 88.3 \\
\enddata
\tablenotetext{a}{Heliocentric JD of mid-integration minus 2,450,000.}
\end{deluxetable}

\begin{deluxetable}{ll}
\tablewidth{0pt}
\tabletypesize{\footnotesize}
\tablecaption{PQ And system parameters\label{tab:parms}}
\tablehead{
\colhead{Parameter} & \colhead{Value} 
}
\startdata
\cutinhead{Observed parameters}
RA (J2000) & +02:29:29.54 \\
DEC (J2000) & +40:02:39.40 \\
b & -19.06$^{\circ}$ \\
E(B-V) & 0.06 \citep{SFD98} \\
Outbursts & 23 August 1938 \citep{richter90} \\
          &  7 March 1967 \citep{richter90} \\
          & 21 March 1988 \citep{iauc4570} \\
Recurrence Timescale & 25$\pm$4 years \\
Maximum magnitude & 10 (visual) \citep{iauc4570} \\
Minimum magnitude & 19.0 (V band) \citep{DS93} \\
\cutinhead{H$\alpha$ emission line analysis} 
Orbital period & 1.7$\pm$0.1 hours \\
Systemic velocity & 46.0$\pm$3.5 km s$^{-1}$ (Equ. 1) \\
Semiamplitude & 135.7$\pm$13.7 km s$^{-1}$ (Equ. 1) \\
Superior conjunction & 2,452,898.902 HJD (Equ. 2)\\
A & 0.95$\pm$0.01 (Equ. 2)\\
B & 0.25$\pm$0.01 (Equ. 2)\\
\cutinhead{{\tt PHOENIX} WD analysis} 
T$_{eff}$ & 12,000$\pm$1,000 K \\
log(g) & 7.7$\pm$0.3 \\
Z & 10$^{-2}$ Z$_\odot$ \\
Distance & 330$\pm$50 pc \\
M$_1$ & 0.47$\pm$0.13 M$_\odot$ \\
\enddata
\end{deluxetable}


\begin{thebibliography}{}

\bibitem[Barman et al.(2000)]{Barman00} Barman, T.~S.,
Hauschildt, P.~H., Short, C.~I, \& Baron, E.,\ 2000, \apj, 537, 946

\bibitem[Bergeron et al.(1995)]{B95} Bergeron, P., 
Wesemael, F., Lamontagne, R., Fontaine, G., Saffer, R.~A., \& Allard, 
N.~F.\ 1995, \apj, 449, 258

\bibitem[Downes \& Shara(1993)]{DS93} Downes R.A., \& Shara M.,
1993, PASP, 105, 127

\bibitem[Driebe et al.(1998)]{Driebe98} Driebe, T., Sch\"onberner, D., 
Bl\"ocker T., \& Herwig, F., \ 1998, \aap, 339, 123

\bibitem[Hoard et al.(2002)]{2MASS} 
Hoard, D.~W., Wachter, S., Clark, L.~L., \& Bowers, T.~P.\ 2002, \apj, 565, 
511 

\bibitem[Howell, Szkody, \& Cannizzo(1995)]{HSC95} Howell, 
S.~B., Szkody, P., \& Cannizzo, J.~K.\ 1995, \apj, 439, 337 

\bibitem[Horne \& Baliunas(1986)]{HB86} Horne, J.~H.~\& 
Baliunas, S.~L.\ 1986, \apj, 302, 757

\bibitem[Hurst et al.(1988a)]{iauc4570} 
Hurst, G.~M., McAdam, D., Mobberley, M., \& James, N.\ 1988, \iaucirc, 
4570, 2 

\bibitem[Hurst et al.(1988b)]{iauc4577} Hurst, G.~M., Young, A.,
Manning, B., Mobberley, M., Oates, M., Boattini, A., \& Scovil, C.\ 1988,
\iaucirc, 4577, 3

\bibitem[Hurst \& Young(1988)]{iauc4579} Hurst, G.~M.~\& Young, 
A.\ 1988, \iaucirc, 4579, 1 




\bibitem[Richter(1990)]{richter90} Richter, G.~A.\ 1990, 
Informational Bulletin on Variable Stars, 3546, 1 

\bibitem[Scargle(1982)]{Scar82} Scargle, J.~D.\ 1982, \apj, 263, 835

\bibitem[Schlegel, Finkbeiner, \& Davis(1998)]{SFD98} 
Schlegel, D.~J., Finkbeiner, D.~P., \& Davis, M.\ 1998, \apj, 500, 525 

\bibitem[Shafter(1985)]{shaft} Shafter, A. W. 1985, in Cataclysmic 
Variable and Low Mass X-Ray Binaries, ed. D. Q. Lamb \& J. Patterson 
(Dordrecht: Reidel), 355

\bibitem[Sproats, Howell, \& Mason(1996)]{SHM96} Sproats, 
L.~N., Howell, S.~B., \& Mason, K.~O.\ 1996, \mnras, 282, 1211 

\bibitem[Szkody et al.(2002)]{P1} Szkody, P., et al. 2002, \aj, 123, 
430 (Paper 1)

\bibitem[van Zyl et al.(2004)]{van04} van Zyl. L. et al. 2004, MNRAS, 350, 307

\bibitem[Wade \& Hamilton(1988)]{iauc4629} Wade, R.~A.~\& 
Hamilton, D.\ 1988, \iaucirc, 4629, 1 

\bibitem[Warner(1995)]{War95} Warner, B., 1995, in Cataclysmic Variable
Stars, (Cambridge University Press: Cambridge)

\bibitem[Winget et al.(1991)]{Win91} Winget, D.~E., et al.\ 
1991, \apj, 378, 326 

\bibitem[Warner \& Woudt(2004)]{WW04b} Warner, B. \&
Woudt, P. A. 2004, in ASP Conf Ser. Variable Stars in the Local
Group, eds. D. Kurtz \& Karen Pollard, in press

\bibitem[Woudt \& Warner(2004)]{WW04a} Woudt, P.~A.~\& 
Warner, B.\ 2004, \mnras, 348, 599

\end{thebibliography}
\end{document}